# Radiation Tests for a Single-GEM Loaded Gaseous Detector

Kyong Sei Lee[1]*, Byungsik Hong[1], Sang Yeol Kim[2], Sung Keun Park[1]

[1]*Korea University, Seoul, Korea*

[2]*NoticeKorea, Anyang, Korea*

We report on the systematic study of a single-gas-electron-multiplication (GEM) loaded gaseous detector developed for precision measurements of high-energy particle beams and dose-verification measurements. In the present study, a 256-channel prototype detector with an active area of 16×16 cm$^2$, operated in a continuous current-integration-mode signal-processing method, was manufactured and tested with x rays emitted from a 70-kV x-ray generator and 43-MeV protons provided by the MC50 proton cyclotron at the Korea Institute of Radiological and Medical Science (KIRAMS). The amplified detector response was measured for the x rays with an intensity of about $5 \times 10^6$ Hz cm$^{-2}$. The linearity of the detector response to the particle flux was examined and validated by using 43-MeV proton beams. The non-uniform development of the amplification for the gas electrons in space was corrected by applying proper calibration to the channel responses of the measured beam-profile data. We concluded from the radiation tests that the detector developed in the present study will allow us to perform quality measurements of various high-energy particle beams and to apply the technology to dose verification measurements in particle therapy.

Keywords: Gas electron multiplier, Gaseous ionization detector, Particle beam, Dose verification

## I. INTRODUCTION

Ionization detectors with a thin material thickness are suitable for measurements of high-energy particle beam profiles and of precision dose verification in particle therapy. Instead of the conventional counting mode in which maximum affordable counting rate per detector channel is restricted to a level of about $10^8$ Hz, we adapt a continuous current-integration mode for the signal process [1-4] that allows a wider measurable range of particle flux.

The magnitude of amplification for gas electrons by using a GEM electrode required for quality measurements for particle beams with fluxes of lower than $10^6$ Hz cm$^{-2}$ is about 20. On the other hand, the gain can be closely adjusted to 1 for particle fluxes of higher than $10^8$ Hz cm$^{-2}$, whose dose rate is in a range of particle therapy. With the assistance of the single-step GEM, the minimum and maximum particle fluxes guaranteeing quality measurements with a statistical fluctuation of a few % in the channel response are $10^5$ and $10^{10}$ Hz cm$^{-2}$, respectively.

To be a proper detector for the beam measurements, small material thickness is essential to minimize the influence of the detector material to the incident beams, *i.e.*, the energy loss due to ionization and the broadening of the beam emittance due to multiple Coulomb scatterings [5]. While the effect of multiple Coulomb scatterings rapidly decreases as the energy of hadrons increases, the ionization energy loss remains a saturated value of 2.0 MeV g$^{-1}$ cm$^2$. Therefore, the detector with a material thickness less than 0.1 g cm$^{-2}$ and a scattering length of about $5 \times 10^{-3}$ $X_0$ [6] is, therefore, the relevant choice for the measurements for proton and heavy-ion beams with energies of about 100 AMeV.

The structure of the detector equipped with a GEM foil is described in Section II. In Section III, we briefly describe the dedicated electronics for the signal process based on the continuous current-integration mode and the DAQ operated with a maximum data-transfer speed of 35 kHz. The test results for x rays emitted from a 70 kV x-ray generator and for 43-MeV protons provided by the MC50 proton cyclotron at the Korea Institute of Radiological and Medical Science (KIRAMS) are described in Sections IV and V, respectively. Finally, the conclusions for the detector performances and potential applications to large-scale x-ray inspections and dose-verification measurements in particle therapy are discussed in Section VI.

## II. DETECTOR AND ELECTRONICS FOR THE SIGNAL PROCESS

The structure of a single-GEM loaded detector is illustrated with the schematic diagram Fig. 1. The thickness for drift and induction regions was adjusted to 3.2 mm. As shown in Fig. 1, a resister chain composed of two 4.0 and 2.0 MΩ resistors provides electrical potentials to the drift and the induction regions, and to the GEM electrode with a ratio of 2:1:2.

The GEM foil was manufactured with a 50-μm thick Kapton film double-side coated with 5-μm thick copper layers. The diameter of GEM holes and the two-dimentional spacing are 50 and 140 μm, respectively. The GEM foil was tightly attached on a 1.6-mm thick printed circuit board (PCB) with adhesive glue without allowing sagging. The cathode plane was manufactured with a 25-μm thick aluminized polyester film in the same way as the GEM foil. The electric potential to be applied across the GEM electrode ranges from 300 to 440 V, when a gas mixture of 70% Ar + 30% $CO_2$ is used for the detector operation.

The signal plate manufactured with a 200-μm-thick PCB is composed of 1.25-mm pitch strips and pad arrays printed on the top layer of the PCB plate, as illustrated in Fig. 2. The strips and the pad arrays, 128 of each, were assigned to measure the separate detector responses in the vertical ($y$) and horizontal ($x$) directions, respectively. The 100-μm-wide traces required for the electrical connections for the pad arrays were printed on the bottom layer. The estimated total material thickness of a single-GEM loaded detector is 0.077 g cm$^{-2}$.

The 256-channel electronics for the signal process is composed of four 64-channel charge integrators, eight amplifiers, an eight-channel analog-digital-converter (ADC) processor, a field-programmable-gate-array (FPGA) digital processor, and a USB3 interface processor. The details for the signal processing electronics are described in the previously report [7].

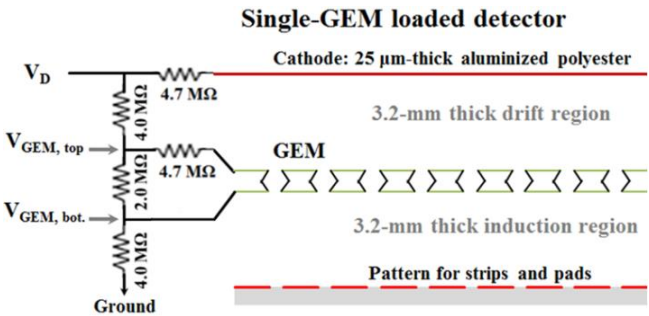

Fig. 1. Schematic diagram for the single GEM-loaded detector.

III. TEST OF THE SINGLE-GEM-LOADED DETECTOR WITH X-RAYS

The single-GEM-loaded detector was place at a distance of 10 cm from the exit window of a 70-kV x-ray generator. The gas for the detector operation is a mixture of 70% Ar + 30% $CO_2$. The signal rate of x rays with a mean energy of about 20 keV emitted from the x-ray generator detected near the central region of detection (in the full width at half maximum; FWHM) was expected to be about $5\times10^6$ Hz cm$^{-2}$. The x ray signals collected in the strips and pad arrays of the detectors are integrated every 114.4 μs, and converted to the integrated charge values (channel responses). The maximum sensitivity of the radiation-induced current per channel was adjusted to 96 nA.

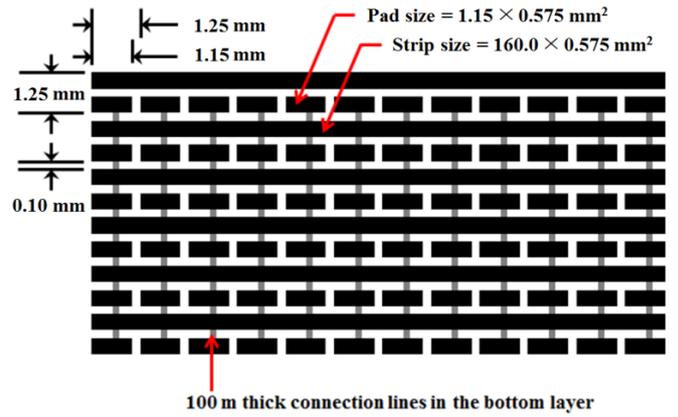

Fig. 2. Copper pattern of 1.25-mm-pitch strips and pad arrays printed on the top side of a 200-μm thick PCB for the single-GEM-loaded detector. The 100-μm-wide traces required for the electrical connections for the pads were printed on the bottom layer of the PCB.

Figure 3 shows total detector response (charge) measured for 4.9 s as a function of the voltage applied to the GEM electrode ($V_{GEM}$). The induced charges of about 65 fC measured at voltages lower than 150 V are mainly due to the contribution of photoelectrons induced in the gas in the induction region, in the GEM foil, and in the signal plate. The ratio of the total detector response at $V_{GEM} = 420$ V to that at $V_{GEM} = 20$ V was measured as 22.5.

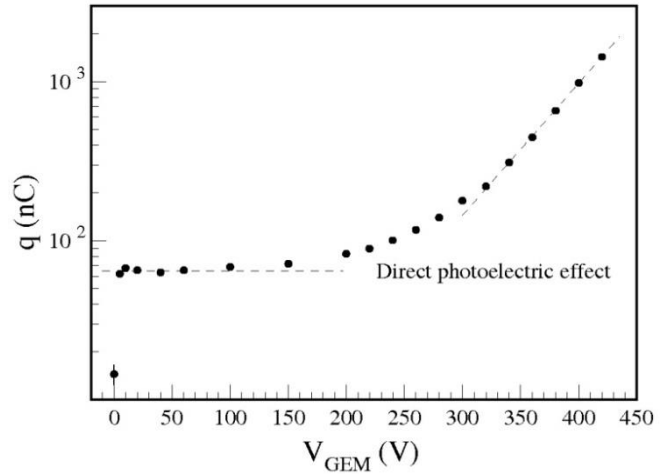

Fig. 3. Total detector response measured as a function of the voltage applied to the GEM electrode.

Fig. 4 shows spatial distributions of the detector responses for the x rays in $x$ (top) and $y$ (bottom) directions, measured at $V_{GEM} = 20$, 340, 380, and 420 V for 4.9 s. The mean ratios of the increment in the voltage interval between 340 and 420 V, measured for the $x$- and $y$-direction responses were, 5.5 and 5.4, respectively.

As shown in Fig. 4, the non-uniform channel responses resulted from the spatially inhomogeneous gain and transparency for the gas electrons through the GEM holes. In order to improve the precision in the measurements, proper corrections for the non-uniform channel responses are necessary. The correction procedure and the results for proton-beam data are discussed in Section V.

Figure 5 shows distributions of charges induced in channel 64 of $x$ ($q_{x64}$, top) and in channel 64 of $y$ ($q_{y64}$, bottom) for 114.4 µs when $V_{GEM}$ was set to 20, 260, 300, 340, 380, 420 V (from left to right, respectively). The ratios of the mean charges at $V_{GEM}$ = 420 V to that at $V_{GEM}$ = 20 V for $q_{x64}$ and $q_{y64}$, were 22.7 and 22.2 respectively.

Figure 6 shows time responses of channel 60 of $x$ (top) and channel 60 of $y$ (bottom) for x rays at $V_{GEM}$ = 420 V. The irradiation of x rays was initiated about 2 s after the DAQ start. $V_{GEM}$ = 420 V was the highest voltage for an operation without discharges when operated with the gas mixture of 70% Ar + 30% $CO_2$. Discharges ruin the time spectra of the channel responses because of recording overflows in the data. The maximum $V_{GEM}$ for practically reliable measurements without being hampered by discharges was 420 V, where the expected gain for the gas electrons is about 35.

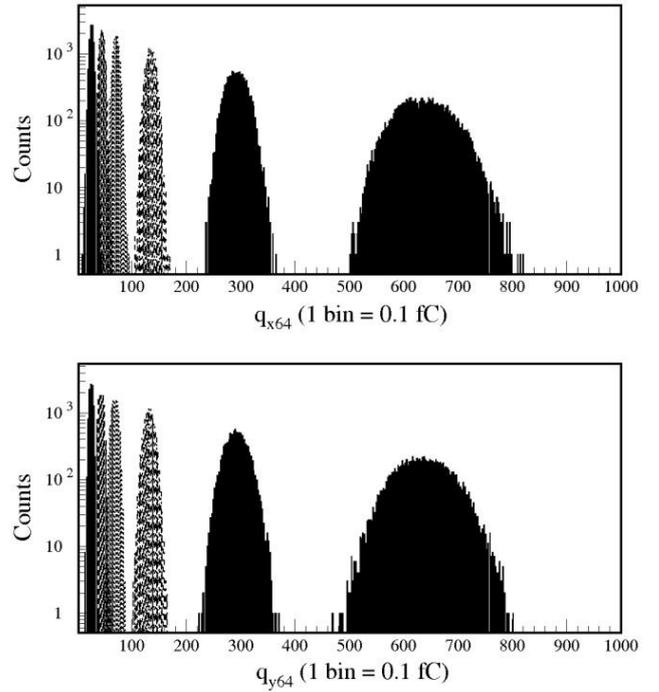

Fig. 5. Distributions of charges induced in channel 64 of $x$ ($q_{x64}$, top) and in channel 64 of $y$ ($q_{y64}$, bottom) for 114.4 µs when $V_{GEM}$ was set to 20, 260, 300, 340, 380, 420 V (from the left to the right, respectively).

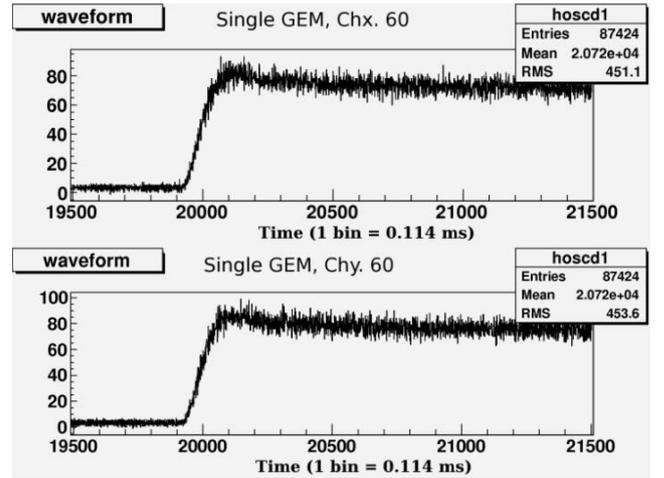

Fig. 6. Time responses of channel 60 of $x$ (top) and channel 60 of $y$ (bottom) for x rays at $V_{GEM}$ = 420 V.

## VI. Test of the single-GEM amplification with 43-MeV proton beams

The single-GEM-loaded detector was tested with 45-MeV proton beams provided by the MC50 cyclotron at the Korea Institute of Radiological and Medical Science (KIRAMS). Firstly, the detector was installed at a distance of 172 cm from the vacuum exit of the proton beam to examine the single-step GEM amplification and the spatial uniformity.

Because of energy losses in a 0.5-mm-thick aluminum vacuum window and in the air lying between the beam exit and the detectors, the actual most probable energy of the protons delivered to the detector was expected to be 43.0 MeV. The mean flux of a Gaussian-shape proton beam of 0.5 nA with a full-width-of-half maximum (FWHM) of 4 cm was approximately $2.5 \times 10^8$ Hz cm$^{-2}$. The gas mixture for the detector operation is the same for the x-ray test. The maximum sensitivity of the radiation-induced current per channel was adjusted to 384 nA.

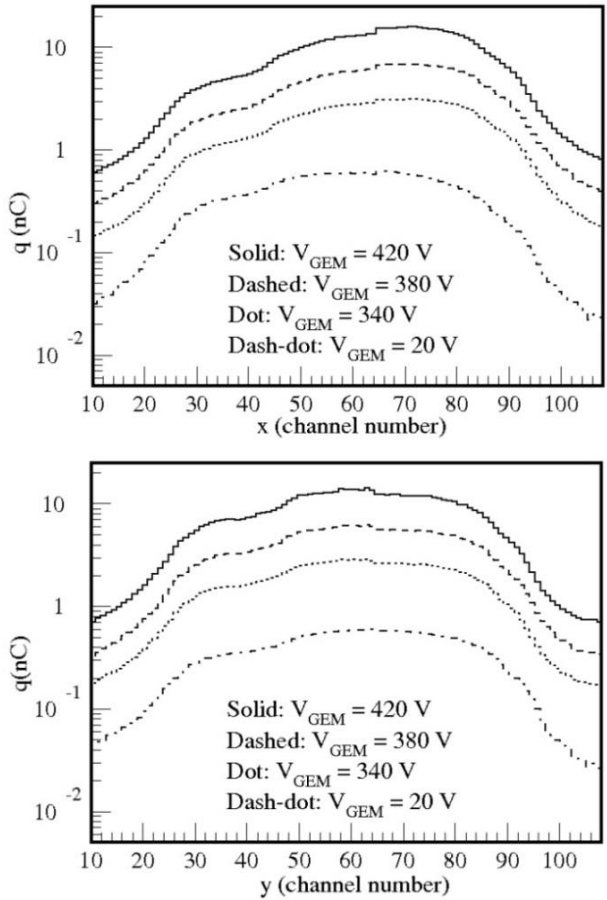

Fig. 4. Spatial distributions of detector responses for the x rays in $x$ (top) and $y$ (bottom) directions, measured at $V_{GEM}$ = 20, 340, 380, and 420 V for 4.9 s.

Figure 7 shows distributions of channel responses for the proton beam in $x$ (top) and $y$ (bottom) directions, measured at $V_{GEM}$ = 120 (dash-dotted), 360 (dot), 400 (dashed), 440 V (solid) for 5.0 s, respectively. The emittances of the beam (defined as standard deviations), measured at the detector position was 1.70 ($x$) and 1.73 cm ($y$), respectively. Ratios of the channel responses at $V_{GEM}$ = 300 (dash-dotted), 360 (dot), 400 (dashed), 440 V (solid) to those at $V_{GEM}$ = 120 V in $x$ and $y$ directions, calculated for the region near the beam center, are shown in the left and right figures in Fig. 8, respectively. At $V_{GEM}$ = 120, the spatially inhomogeneous GEM amplification is expected to be negligible as well as the magnitude of the gain.

smoothed functions of the data at $V_{GEM}$ = 400 V to those at $V_{GEM}$ = 120 V where the shape of the beam profile was preserved without deformation. Then, the two calibration functions, each for $x$ and $y$, were applied to the other data measured at the neighboring voltages. Figure 9 shows the calibrated channel responses in $x$ (top) and $y$ (bottom) directions measured at 360 (dash-dotted), 380 (dotted), 420 (dashed), and 440 V (solid). The central regions (within 2σ) of the calibrated beam-profile data measured at the wide range of $V_{GEM}$ are fairly well agreed with the expected Gaussians. The minimum particle flux to be measured by the present single-GEM-loaded detector is about $10^5$ Hz cm$^{-2}$, when we adjust the magnitude of the gain to about 20 (obtained at $V_{GEM}$ = 400) and apply the proper calibrations for the GEM-amplified channel responses.

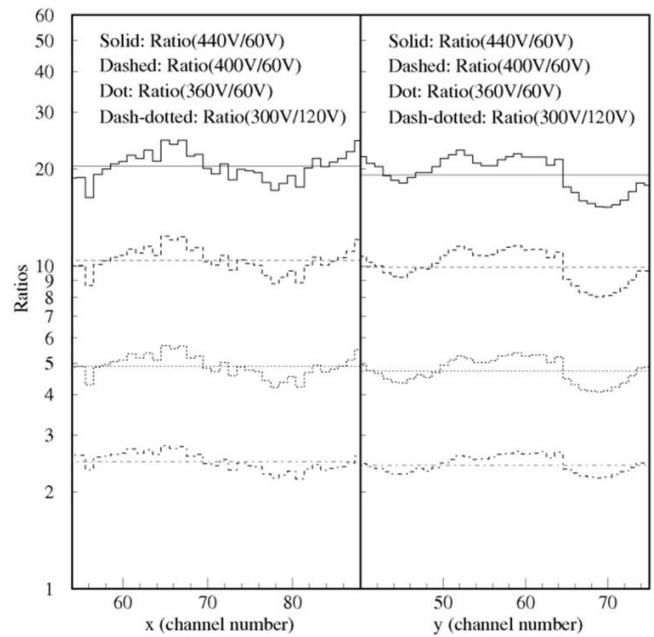

Fig. 8. Ratios of the channel responses at $V_{GEM}$ = 300 (dash-dotted), 360 (dot), 400 (dashed), 440 V (solid) to those at $V_{GEM}$ = 120 V in $x$ (left) and $y$ (right) directions calculated for the region near the beam center.

Two thin plane detectors composed of just a 3.2-mm-thick induction region and, thus, operated in a pure ionization mode were tested together to demonstrate the detection characteristics of the single-GEM-loaded detector. In order to reconcile the dynamic ranges of detections for two different type detectors, $V_{GEM}$ for the single-GEM-loaded detector was set at 240 V, where the gain through the GEM is expected to be approximately 2. The voltage applied to the induction region of the thin plane detectors was 800 V.

The single-GEM-loaded and two thin plane detectors were installed at distances of 130.0, 133.6, and 137.2 cm from the beam vacuum exit, respectively.

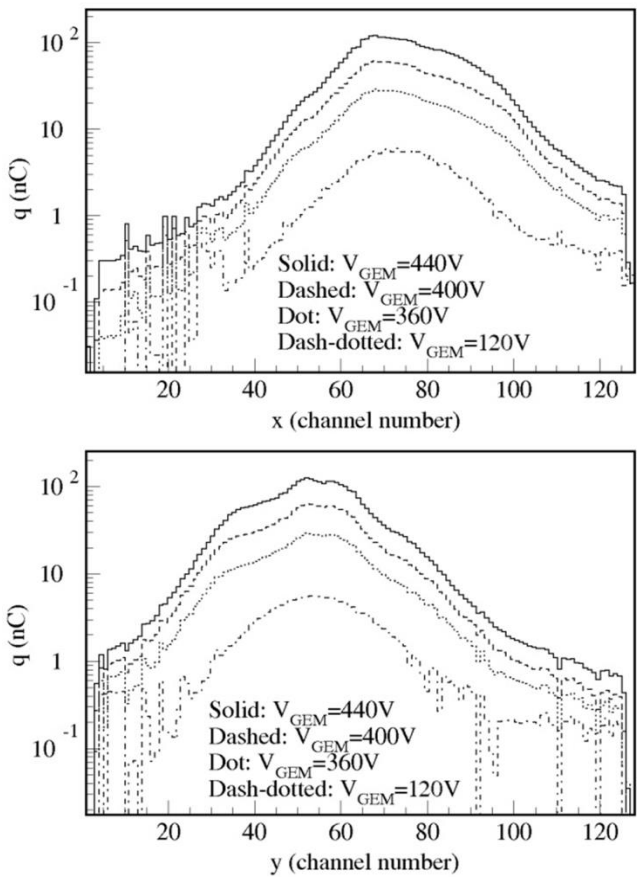

Fig. 7. Distributions of channel responses for the proton beam in $x$ (left) and $y$ (right) directions, measured at $V_{GEM}$ = 120 (dash-dotted), 360 (dot), 400 (dashed), 440 V (solid) for 5.0 s.

Two calibration functions for the non-uniform channel responses were obtained from the ratios of the The position of the first and the second thin plane detectors in the beam depth were estimated to be 0.658 ± 0.015 and 1.300 ± 0.030 g cm$^{-2}$, where the specific energy of the 43-MeV protons predicted by the GEANT4 simulation were 17.4 and 27.5 MeV g$^{-1}$ cm$^2$, respectively.

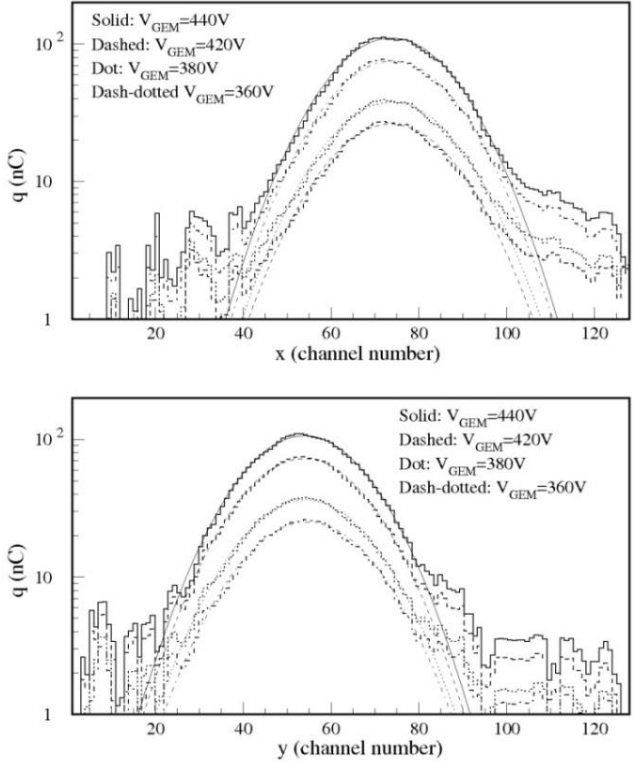

Fig. 9. Calibrated channel responses in $x$ (left) and $y$ (right) directions measured at 360 (dash-dotted), 380 (dotted), 420 (dashed), and 440 V (solid).

Figure 10 shows the detector responses in $x$ (left) and $y$ (right) directions for 0.5 (dash-dotted), 1.0 (dot), 2.0 (dashed), 3.0 (solid) nA proton beam measured the single-GEM (top), the first thin plane (middle), and the second thin plane detectors (bottom). The detector responses were not clearly proportional to the nominal beam intensities due to the poor accuracy of the beam setup using a Faraday cup. However, as shown in Fig. 10, the ratios of the detector responses for 1.0-, 2.0-, and 3.0-nA beams to those for the 0.5-nA beam measured by the three detectors fairly coincide. The mean ratio of the total detector responses measured by the first and the second thin plane detectors, averaged over all the beam data, was valued as 1.555 ± 0.055, which is well agreed with the ratio of the specific energy losses at the given depths, 1.580, predicted by the GEANT4 simulations.

In Fig. 11, total detector responses of the single-GEM-loaded detector for 0.5-nA 43-MeV protons (full circles) measured as a function of $V_{GEM}$ are compared with those previously measured for the x rays (open circles). The estimated mean flux of the 0.5-nA proton beam with a FWHM of 4 cm was $2.5 \times 10^8$ Hz cm$^{-2}$, which was estimated to be about 50 times higher than that for previously measured x-ray data (~ $5 \times 10^6$ Hz cm$^{-2}$). As shown in Fig. 11, the trend of the exponential growth of the gain measured for the protons is less steep than that measured for the x rays because of the higher particle rate. However, the linearity of the detector response to the particle flux was clearly preserved at the particle flux of $2.5 \times 10^8$ Hz cm$^{-2}$, which is still below the expected rate-capability limit (~ $10^9$ Hz cm$^{-2}$) for the single-step GEM amplification detection.

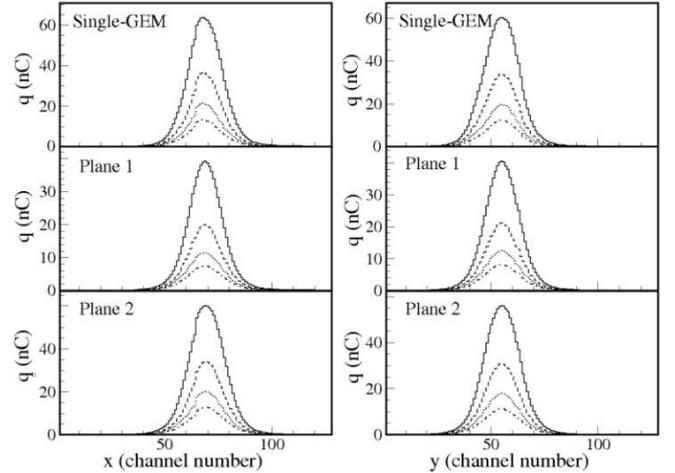

Fig. 10. Detector responses in $x$ (left) and $y$ (right) directions for 0.5 (dash-dotted), 1.0 (dot), 2.0 (dashed), 3.0 (solid) nA proton beam measured by the single-GEM (top), the first thin plane (middle), and the second thin plane detectors (bottom).

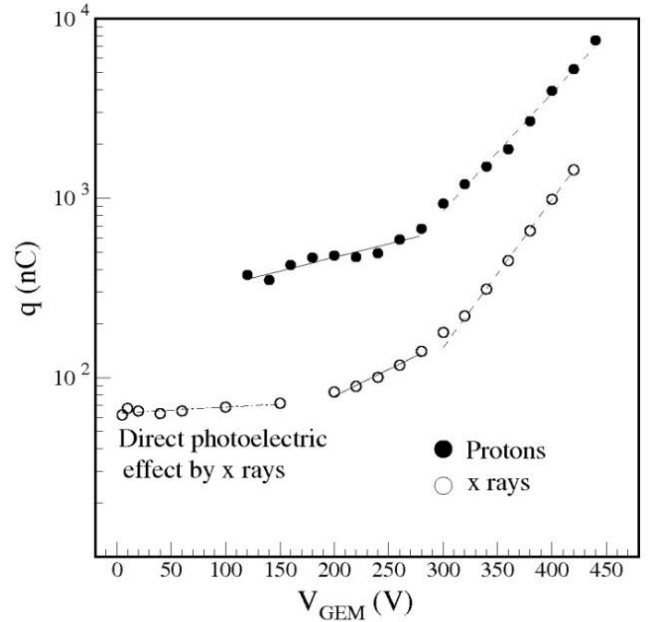

Fig. 11. Total detector responses of the single-GEM-loaded detector for 0.5-nA 43-MeV protons (full circles) and x rays (open circles) measured as a function of $V_{GEM}$.

The position resolution for the single-GEM-loaded detector was measured by placing a 20-mm thick polymethyl-methacrylate (PMMA) collimator with 2-mm wide holes at 2 cm from the detector window. The single-GEM detector was installed at 50 cm from the vacuum exit of a 10-nA proton beam. Figure 12 shows the beam profiles in *x* (left) and *y* (right) directions, collimated by a collimator with a single hole (top) and by one with a 2 × 2 hole matrix with a 20-mm spacing in both direction (bottom). The mean standard deviations of the narrow-beam profiles in the *x* and the *y* directions were valued as 0.743 and 0.604 mm, respectively. The intrinsic position uncertainty of the protons passing a 2-mm hole is expected to be $\frac{\pi}{4} \times \frac{2\,mm}{\sqrt{12}}$ = 0.453 mm. By quadratically subtracting the position uncertainty of 0.453 mm, the position resolutions (standard deviations) obtained for the *x* (pad arrays) and for *y* (strips) directions are valued as 0.588 and 0.400 mm, respectively. The intrinsic position uncertainty of measuring a particle track with 1.25-mm-pitch strips is 0.361 mm. Therefore, the dispersion of the drifting gas electrons in the thin single-GEM-loaded detector affecting on the position resolution is quite insignificant.

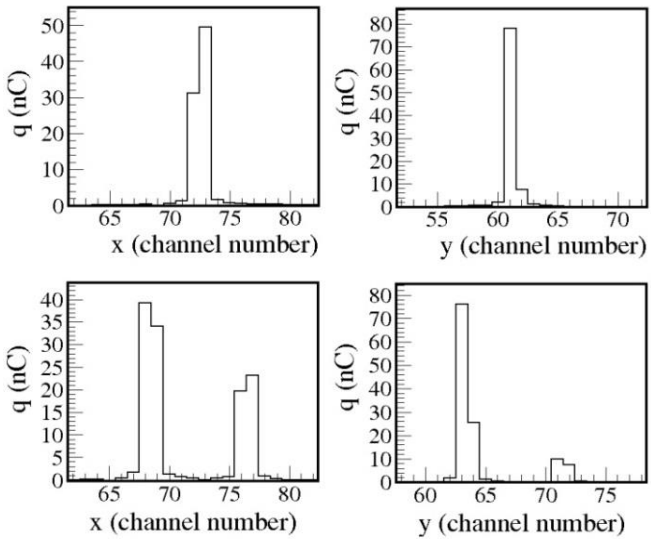

Fig. 12. Beam profiles in *x* (left) and *y* (right) directions, collimated by a collimator with a single hole (top) and by one with a 2 × 2 hole matrix with a 20-mm spacing in both direction (bottom).

Figure 13 shows time profiles for a 3.0-nA proton beam measured by channels 65 of *x* of the single-GEM (top), the first thin plane (middle), and the second thin plane (bottom) detectors. The voltages applied to the single-GEM-loaded and the two thin plane detectors are same with those for the measurements shown in Fig. 10. In Fig. 13, the microscopic pattern of the cyclotron beam in time is clearly observed in the data. Furthermore, it was confirmed that the phase of the beam structure observed in the three detectors well coincide because the channel 65 of *x* for all three detectors lies at the same emittance position. The difference of the magnitudes in the time spectra in the middle and the bottom panels respects the ratio of the detector responses measured by the two thin plane detectors at the given depth positions, *i.e.*, 1.555, as explained in Fig. 10.

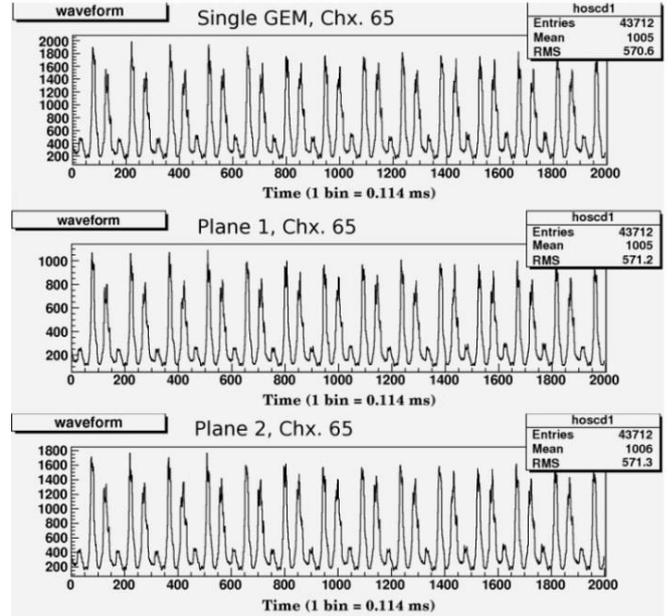

Fig. 13. Time profiles for a 3.0-nA proton beam measured by channel *x*65 of the single-GEM (top), the first thin plane (middle), and second thin plane (bottom) detectors.

V. CONCLUSIONS

The single-GEM-loaded detector has been constructed and tested for precision measurements of high-energy hadron beams. The detector characteristics have been examined with x rays and 43-MeV protons with a maximum particle flux of about $10^9$ Hz cm$^{-2}$. The electronics for the signal process and the DAQ with a maximum data-transfer speed of 35 kHz was designed and manufactured for precision time-dependent measurements of hadron-beam profiles. Here, the conclusions for the present detector R&D are summarized as follows:

(1) The maximum gain obtained by the single-step GEM amplification without allowing discharge was 35 measured at $V_{GEM}$ = 420 V. The maximum gain of about 20 is required to perform statistically qualified measurements for particle-beam profiles with fluxes of less than $10^6$ Hz cm$^{-2}$.

(2) The non-uniform GEM-amplified channel responses was properly calibrated by applying the calibration functions obtained by the ratios of the channel response functions of the sample data measured at $V_{GEM}$ of 120 (unamplified) and 400 V.

(3) The quantitative accuracies of the detector responses in space (Fig. 10) and time (Fig. 14) measured by the single-GEM-loaded detector were confirmed by the comparisons with those measured by the two thin plane detectors prepared for the reference detectors.

(4) To enhance the detector rate capability for high particle rates, the gain of the GEM was reduced to about 2. Then, the linearity of the detector response to the beam was confirmed with the maximum proton fluxes of about $10^{10}$ Hz cm$^{-2}$, which is a factor of a thousand higher than the rate capability for the typical triple-GEM detectors [7, 8].

(5) The intrinsic position resolution for the single-GEM-loaded detector was valued as about 0.5 mm. The dispersion of the drifting gas electrons affecting on the position resolution was insignificant.

The one of the prospective applications of the detector technology developed in the present research will be to dose verification measurements in proton [9, 10] or carbon therapy [11-14]. The small detector thicknesses of the single-GEM-loaded and the thin plane detectors are also advantageous for developing a multilayer detector system, which would dramatically reduce the time and the labor required for executing the dose verification procedure.

The other potential application is to large-scale high-energy x-ray inspections for cargo containers and vehicles. The uniform thickness of the cathode material in the planar-type gaseous detectors to produce Compton electrons will guarantee fairly uniform channel responses to gamma and x rays. Furthermore, a high resolution for scanned images of better than 300 μm can be reliably achieved by adapting a high-density strip or pad readout with a pitch of narrower than 1 mm, which would be practically difficult to achieve with a detector system composed of scintillator crystals and silicon pindiode arrays.

ACKNOWLEDGMENTS

This study was supported by the Korea University Grant (research fellow program) and the National Research Foundation of Korea (Grant Numbers NRF-2013R1A1A2060257 and NRF-2013M2B2A9A-03050128).